\documentclass[aps,pre,twocolumn,showpacs,groupedaddress]{revtex4}  
\usepackage{graphicx}  
\usepackage{dcolumn}   
\usepackage{bm}        
\usepackage{amssymb}   
\usepackage{float}
\usepackage{enumerate}
\begin{document}
\hspace{5.2in}
\title{Imperfect synchronized states and chimera states in two interacting populations of nonlocally coupled Stuart-Landau oscillators}
\author{K. Premalatha$^{1}$, V. K. Chandrasekar$^{2}$, M. Senthilvelan$^{1}$, M. Lakshmanan$^{1}$}
\address{$^1$Centre for Nonlinear Dynamics, School of Physics, Bharathidasan University, Tiruchirappalli - 620 024, Tamil Nadu, India.\\
$^2$Centre for Nonlinear Science \& Engineering, School of Electrical \& Electronics Engineering, SASTRA University, Thanjavur -613 401,Tamilnadu, India.}
\begin{abstract}
 We investigate the emergence of different kinds of imperfect synchronized states and chimera states in two interacting populations of nonlocally coupled Stuart-Landau oscillators.  We find that the complete synchronization in population-I and existence of solitary oscillators which escape from the synchronized group in population-II lead to imperfect synchronized states for sufficiently small values of nonisochronicity parameter.  Interestingly, on increasing the strength of this parameter further there occurs an onset of mixed imperfect synchronized states where the solitary oscillators occur from both the populations.  Synchronized oscillators from both the populations are locked to a common average frequency.  In both the cases of imperfect synchronized states synchronized oscillators exhibit periodic motion while the solitary oscillators are quasi-periodic in nature.  In this region, for spatially prepared initial conditions, we can observe the mixed chimera states where the coexistence of synchronized and desynchronized oscillations occur from both the populations.  On the other hand, imperfect synchronized states are not always stable, and they can drift aperiodically due to instability caused by an increase of nonisochronicity parameter.   We observe that these states are robust to the introduction of frequency mismatch between the two populations.
\end{abstract}

\maketitle

\section{Introduction} 
\par The nature of the dynamics of networks of coupled oscillators and their complex behaviors have been studied for many years \cite{1}. Nowadays considerable interest is shown on analyzing the questions regarding the emergence of chimera states.  These states are characterized by the coexistence of both synchronized and desynchronized behaviors of coupled identical oscillators.  Such a remarkable phenomenon was initially found in nonlocally coupled identical oscillators \cite{2,5,ma}.  It has been subsequently studied in globally coupled oscillator networks \cite{7}, planar oscillators \cite{37a}, heterogeneous networks \cite{36}, oscillators with more than one populations \cite{37t1, 38m1, 38b}, two dimensional map lattices \cite{10} and experimentally in chemical oscillators \cite{34}, an optical system \cite{7a}, electrochemical \cite{34a} and coupled mechanical oscillators \cite{33}.  They have also been identified in certain locally coupled systems \cite{33a,33b} as well. Many theoretical, numerical and experimental investigations deal with a single population consisting of identical oscillators.  However systems under various situations including ensembles of oscillators with more than one population and introduction of coupling asymmetries are less investigated till date.  Owing to the strong resemblance of chimera states with real world applications, investigation around the chimera states is even more important due to the strong relevance of such states with many natural phenomena including unihemispheric sleep of certain mammals and birds where one brain hemisphere appears to be inactive while the other remains active \cite{27}, ventricular fibrillation \cite{27f} (one of the primary causes of sudden cardiac death in humans), blackouts of power grid networks \cite{28p}, social systems  \cite{28s}(organization of coupled populations), neural systems \cite{28n} (firing patterns of neurons, coordinated and uncoordinated brain activity, etc.) and so on.
\par  Moreover another interesting pattern, namely imperfect chimera state is reported in \cite{Imperfect} with coupled pendula and this state is characterized by a certain small number of solitary oscillators (solitary state \cite{34a}) which escapes from the synchronized chimera's cluster (where solitary oscillator represents a single repulsive oscillator splitting up from the fully synchronized group).  Such escaped oscillators oscillate with different average frequencies.  A novel mechanism for the creation of chimera states via the appearance of the solitary states is also reported in Kuramoto model with inertia \cite{39a} and with time delayed feedback oscillators \cite{39b}.  In the present study we aim to investigate different kinds of imperfect synchronized states and chimera states (for spatially prepared initial conditions) in two interacting populations of nonlocally coupled oscillators.  The imperfect synchronized state here is characterized by a certain small number of solitary oscillators exhibiting quasi-periodic oscillations which escapes from the synchronized group. 
\par Taking into account the above facts, we study the dynamics of nonlocally coupled two interacting populations of Stuart-Landau oscillators.  We are interested to investigate how does the nonisochronicity parameter ($c$) affect the emergence of different kinds of imperfect synchronized states and chimera states in such a system with nonlocal coupling.  We find that for given strengths of inter- and intra-population couplings the emergence of imperfect synchronized states for sufficiently smaller values of nonisochronicity parameter ($c$) which means that the synchronized and escaped oscillators from synchronized state exist within population-II while the population-I remains synchronized.  By increasing the strength of this parameter, we find that the synchronized oscillators from both the populations get locked to a common average frequency while the solitary oscillators are distributed with random average frequencies and we term such a state as a mixed imperfect synchronized state.  In addition, synchronized oscillators exhibit periodic motion around the origin, whereas the desynchronized oscillators exhibit quasiperiodic motion but their center of rotation is shifted from the origin.  In this region, for spatially prepared initial conditions, we can observe the coexistence of synchronized and desynchronized oscillations in both the populations, namely mixed chimera states, which is distinct from the results discussed in Ref. \cite{5} where the chimera state represents the complete synchronization in one population while desynchronization occurs among the oscillators in the other population under global coupling.   We can also find that the imperfect synchronized states can drift with time by increasing the parameter $c$.  We also find that these states are robust against an introduction of frequency mismatch between the natural frequencies of the population with significant values of nonisochronicity parameter.   
\par The structure of the paper is organized as follows. In section-II, we introduce the model of two interacting populations of nonlocally coupled Stuart-Landau oscillators and present the different dynamical states including mixed chimera states, imperfect synchronized states, mixed imperfect synchronized states and drifted imperfect synchronized states.  In section-III, we illustrate the robustness of these states for the introduction of frequency mismatch between the two populations.  We summarize our findings in section-IV.  

\section{Study of mixed quasi-periodic solitary and chimera states in two interacting populations of nonlocally coupled Stuart-Landau oscillators}
\subsection{Model}
To appreciate the results mentioned above, we consider a system of nonlocally coupled two populations of Stuart-Landau oscillators which is described by the following set of coupled equations,

{\small \begin{eqnarray}
\dot{z_j}^{(1,2)}=(1+i\omega)z_{j}^{(1,2)}-(1- ic)|z_{j}^{(1,2)}|^2 z_{j}^{(1,2)} \nonumber\\+\frac{\sigma}{2P_1} \sum_{k=j-P_1}^{j+P_1} (z_{k}^{(1,2)}-z_{j}^{(1,2)})+\frac{\eta}{2P_2} \sum_{k=j-P_2}^{j+P_2} (z_{k}^{(2,1)}-z_{j}^{(1,2)}), \nonumber\\j=1,2,...,N,\qquad \qquad
\label{g}
\end{eqnarray}}
where the complex dynamical variables $z_j^{(1,2)}=x_j^{(1,2)}+iy_j^{(1,2)}$, $j=1,2,...,N$, $c$ is the nonisochronicity parameter, $\omega$ is the natural frequency of the oscillators, and $\sigma$ and $\eta$ represent the strengths of the coupling interactions within and between the populations, respectively.   In system (\ref{g}), each oscillator is coupled with $P_1$ oscillators within its group and $P_2$ oscillators with the other group.  Here superfices 1 and 2 for the variables $z_j$ (or equivalently $x_j$ and $y_j$) refer to population-I and population-II, respectively.  Generally communities of oscillatory network consisting of interacting subpopulations are common in many natural systems.  For example, observation of neuronal activity is taken from different regions of the brain which forms a network of interacting subpopulations of the brain \cite{28n}.  Similarly, man made complex networks, namely power grid networks \cite{28p}, social networks \cite{28s}, etc. constitute coupled networks.  In most of the cases, connection between such subpopulations are with finite number of nodes/oscillators in each subpopulation.  In this connection, for simplicity we have chosen the case where $P=P_1=P_2$ in equation (\ref{g}) and the coupling range is $r=\frac{P}{N}$.
\par In our simulations, we choose generally the number of oscillators $N$ to be equal to 100 and in order to solve Eq. (\ref{g}), we use the fourth order Runge-Kutta method with time step 0.01.  We allow $5\times10^{5}$ iteration time steps as transients.  We have verified that the results are independent of the increase in the number of oscillators.  Note that Figs. \ref{f3} and \ref{f3a} below are plotted for $N=500$.  
\begin{figure}[ht!]
\begin{center}
\includegraphics[width=1.0\linewidth]{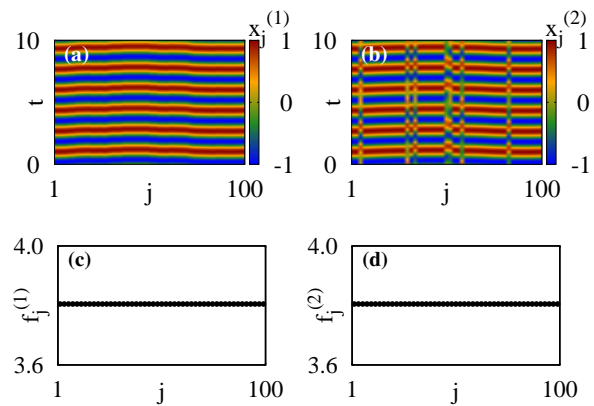}
\end{center}
\vspace{-0.5 cm}
\caption{(Color online) Space-time plots of the variables $x_j^{(1,2)}$ for imperfect synchronized state (a) for population-I and (b) for population-II.  Corresponding oscillator average frequencies of (c) population-I and (d) population-II.  Parameter values: $c=2.3$, $\sigma=0.1$, $\eta=0.25$, $\omega=1.0$ and $r=0.1$.}
\label{f1}
\end{figure}
\begin{figure}[ht!]
\begin{center}
\includegraphics[width=1.0\linewidth]{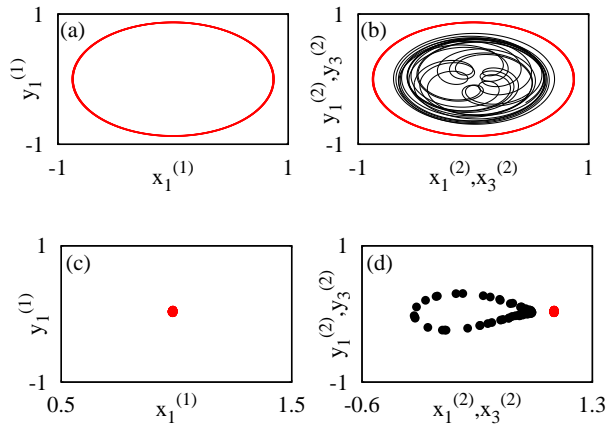}
\end{center}
\vspace{-0.5 cm}
\caption{(Color online) Phase portraits of the oscillators: (a) periodic oscillation of synchronized oscillator $z_1^{(1)}(=x_1^{(1)}+iy_1^{(1)})$. (b) Periodic motion  of the synchronized oscillator $z_1^{(2)}$ and quasi-periodic oscillation of solitary oscillators $z_3^{(2)}$.  Their corresponding Poincar\'{e} surfaces of section: in (c),(d) red/gray dot represents the periodic oscillation of synchronized oscillator and black dots represent the quasi-periodic oscillation of solitary oscillator.  Parameter values: $c=2.3$, $\sigma=0.1$, $\eta=0.25$, $\omega=1.0$ and $r=0.1$.}
\label{fp}
\end{figure}
\begin{figure}[ht!]
\begin{center}
\includegraphics[width=1.0\linewidth]{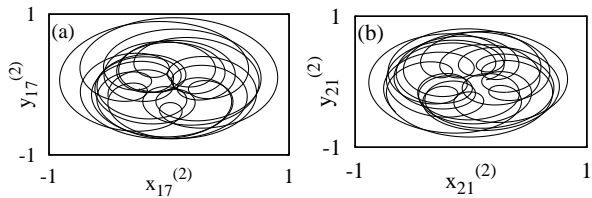}
\end{center}
\vspace{-0.5 cm}
\caption{(a) and (b) Phase portraits of the solitary oscillators $z_{17}^{(1)}$ and $z_{21}^{(2)}$ which show the variation in the centers of rotation.}
\label{p2}
\end{figure}

\subsection{Imperfect synchronized states}
\par To explore the dynamics of the system (\ref{g}), we perform numerical investigations  by considering the natural frequencies of the oscillators as same in both the populations (we will relax this condition later on in our study).  We choose the synchronized initial state to the oscillators in population-I and distribute the initial state of the oscillators in population-II uniformly between $-1$ and $+1$ for the variables $x_j^{(2)}$ and $y_j^{(2)}$ independently.  To start with, we choose the specific choice of $c$ as $c=2.3$ for the fixed coupling range $r(=\frac{P}{N})=0.1$ and natural frequency $\omega=1.0$.  By fixing the value of the coupling parameter $\sigma=0.1$ over a range of $\eta$ we study the dynamical behaviour of the oscillators.  Varying the value of $\eta$, for $\eta=0.25$, we can observe synchronized oscillations in population-I and the existence of solitary oscillators in population-II resulting in the existence of a imperfect synchronized state which is illustrated with space-time plots in Figs. \ref{f1}(a,b).  The imperfect synchronized state is characterized by a certain small number of solitary oscillators exhibiting quasi-periodic oscillations which escapes from the synchronized group.  In Figs. \ref{f1}(c,d), we can observe that the synchronized oscillators in population-I are entrained to a common average frequency and in population-II variations occur only in the amplitudes of the oscillators while the average frequency of the coherent and incoherent oscillators are the same.  Here the incoherent oscillators are non-phase coherent so that we find approximate average frequency of the oscillators by calculating the number of maxima in the time series of a variable in a given time interval $\Delta T$.  It is calculated from the expression $f_j^{(1,2)}= 2\pi \Omega_j^{(1,2)}/\Delta T$, where $j=1,2,3,...N$ and $\Omega_j^{(1,2)}$'s are the number of maxima in the time series $x_j^{(1,2)}$ of the $j^{th}$ oscillator with time interval $\Delta T$ calculated between the integration time step units $5 \times 10^5$ and $1\times 10^6$, after allowing transients of the order of $5 \times 10^5$ time units.  We also observe that synchronized oscillators from both the populations exhibit periodic oscillations while the solitary oscillators from population-II exhibit quasi-periodic oscillations.  Fig. \ref{fp}(a) shows the phase portrait of the synchronized oscillator $z_1^{(1)}$ which shows the periodic motion of this oscillator.  In Fig. \ref{fp}(b), we can observe that the periodic motion of the synchronized oscillator $z_1^{(1)}$ and quasiperiodic motion of the solitary oscillator $z_3^{(1)}$.  This is also confirmed with the Poincar\'{e} surfaces of section corresponding to the above mentioned synchronized oscillator (red/grey dot) and for the solitary oscillator  (black dots) in Figs. \ref{fp} (c) and (d), respectively.   This imperfect synchronized state has similarity with the amplitude chimera state reported in Ref. \cite{39c} where both synchronized and dsynchronized oscillations are periodic in nature. However oscillators from the synchronized group perform oscillations around the origin, whereas for the oscillators in the desynchronized group, the center of rotation is shifted from the origin.  In the present case of imperfect synchronized state, synchronized oscillators are oscillating around the origin and are periodic in nature.  On the other hand, the center of rotation (of orbit with small amplitude) of all the solitary oscillators are shifted from the origin but their motion is quasi-periodic in nature. Such shifts in the centers of rotation of solitary oscillators which is illustrated for two different solitary oscillators labeled as $z_{17}^{(2)}$ and $z_{21}^{(2)}$ in Figs. \ref{p2}(a) and (b).  We also note here the emergence of this type of imperfect synchronized states no longer persists for a sufficiently high value of the nonisochronicity parameter ($c$).  
\begin{figure}[ht!]
\begin{center}
\includegraphics[width=1.0\linewidth]{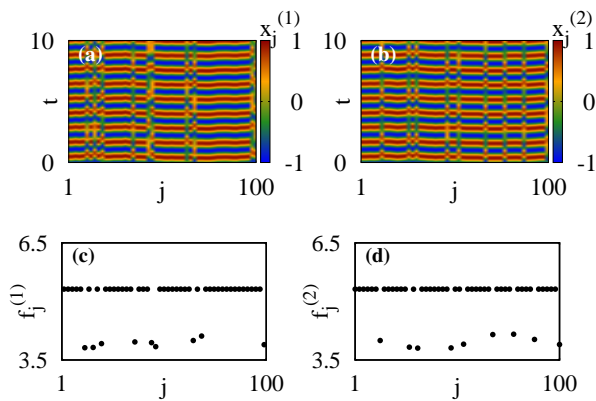}
\end{center}
\vspace{-0.5 cm}
\caption{(Color online) Space-time plots of the variables $x_j^{(1,2)}$ for mixed imperfect synchronized state (a) for population-I and (b) for population-II.  Corresponding oscillator average frequencies of (c) population-I and (d) population-II.  Parameter values: $c=5$, $\sigma=0.1$, $\eta=0.25$, $\omega=1.0$ and $r=0.1$.}
\label{f2}
\end{figure}
\begin{figure}[ht!]
\begin{center}
\includegraphics[width=1.1\linewidth]{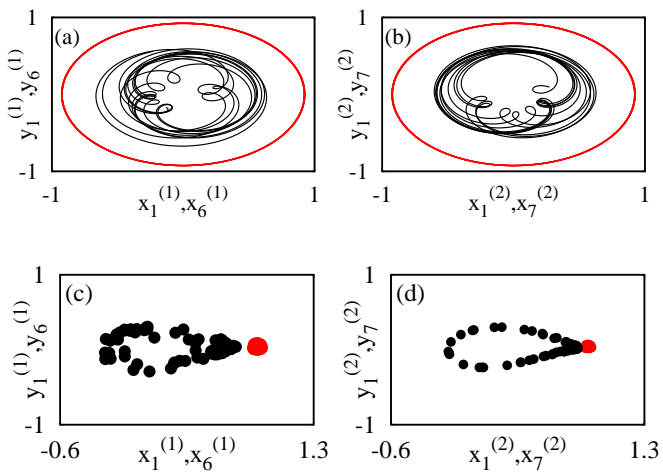}
\end{center}
\vspace{-0.5 cm}
\caption{(Color online)  Phase portraits of the oscillators: (a) periodic oscillation of synchronized oscillator $z_1^{(1)}$ and quasi-periodic oscillation of solitary oscillator $z_6^{(1)}$. (b) Periodic motion  of the synchronized oscillator $z_1^{(2)}$ and quasi-periodic oscillation of solitary oscillator $z_7^{(2)}$.  Their corresponding Poincar\'{e} surfaces of section: in (c), (d) red/gray dot represents the periodic oscillation of synchronized oscillator and black dots represent the quasi-periodic oscillation of solitary oscillator.  Parameter values: $c=5$, $\sigma=0.1$, $\eta=0.25$, $\omega=1.0$ and $r=0.1$.}
\label{fp2}
\end{figure}
\subsection{Mixed imperfect synchronized states}
\par Interestingly, an increase in the strength of $c$ leads to the onset of a new type of imperfect synchronized state.  For the system parameter values $\sigma=0.1$, $\eta=0.25$, and $r=0.1$ with $c=5$, we can observe the solitary oscillators where the synchronized group is having oscillators from both the populations with the same average frequency while the solitary group is having oscillators with random average frequencies.  This state is designated as a mixed imperfect synchronized state and is demonstrated with space-time plots in Figs. \ref{f2}(a,b) and average frequency profiles of the oscillators in Figs. \ref{f2}(c,d).  In the case of mixed imperfect synchronized state, synchronized oscillators labeled as $z_1^{(1)}$ and $z_1^{(2)}$ from both the populations are oscillating periodically (red/grey curve in Figs. \ref{fp2}(a) and (b)).  The deviated oscillators labeled as $z_6^{(1)}$ and $z_7^{(2)}$ (that is solitary oscillators) from both the populations exhibit quasiperiodic motion (black curve in Figs. \ref{fp2}(a) and (b)).  Periodic and quasi-periodic oscillations of the corresponding oscillators are confirmed with the Poincar\'{e} surfaces of section in Figs. \ref{fp2}(c) and (d).  Upon increasing the coupling strength to larger values, the oscillators attain a completely synchronized state which is discussed in the following.\\

\begin{figure}[ht!]
\begin{center}
\includegraphics[width=1.1\linewidth]{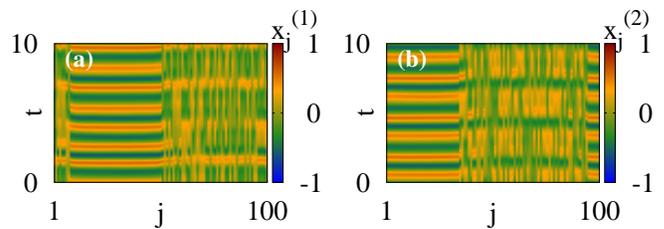}
\end{center}
\vspace{-0.5 cm}
\caption{(Color online) Space-time plots of the variables $x_j^{(1,2)}$ for mixed chimera state for spatially prepared initial conditions (a) population-I and (b) for population-II.  Other parameter values are $c=5$, $\sigma=0.1$, $\eta=0.25$, $\omega=1.0$ and $r=0.1$.}
\label{f2a}
\end{figure}

 Another interesting phenomenon to be noted here is that one can observe the coexistence of regions of synchronized and desynchronized oscillations, namely the chimera state in both the populations.  Hence this state is designated as mixed chimera state which is illustrated in Figs. \ref{f2a}(a), (b).  Here the synchronized oscillators are oscillating periodically but the desynchronized oscillators are quasiperiodic in nature which is different from the state reported in ref. \cite{39c}.  Note that the above type of state can be achieved for only spatially prepared initial conditions for both the populations independently.  If we perturb the system from this initial state of the oscillators, the system enters into the mixed imperfect synchronized states. 
\subsection{Drifted imperfect synchronized states}

\par
\begin{figure}[ht!]
\begin{center}
\includegraphics[width=1.0\linewidth]{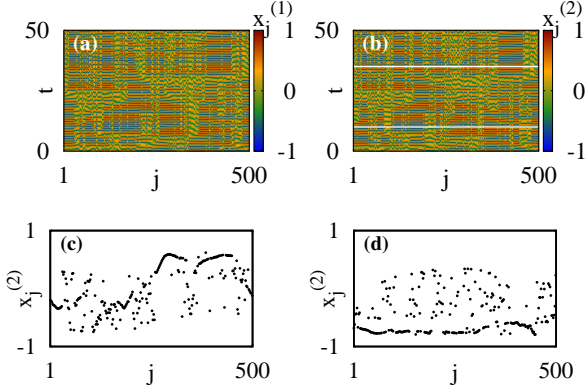}
\end{center}
\vspace{-0.7 cm}
\caption{(Color online) Space-time plots of the variables $x_j^{(1,2)}$ (a) for population-I and (b) for population-II.  Snapshots for the variables $x_j^{(2)}$ of population-II for drifted imperfect  synchronized state (at two different times which is marked by the white line in Fig. \ref{f3}(b)) (c) $t=20$, (b) $t=35$.  The other parameter values are $c=9$, $r=0.1$, $\sigma=0.1$, $\eta=0.15$, $\omega=1.0$.}
\label{f3}
\end{figure}

 We also analyze the stability of the mixed imperfect synchronized states for an increase of the nonisochronicity parameter values.  For the value of $c=9$ with the same system parameter values considered for mixed imperfect synchronized states, we can observe that the above state becomes drifting with time where the synchronized group of oscillators exist for certain time period after which it escapes from synchronized group of oscillators.  This state is designated as drifted imperfect  synchronized state.  Such drifting of solitary oscillators from synchronized group occurs in an aperiodic manner.  We can clearly observe the existence of drifted imperfect synchronized state in Figs. \ref{f3}(a,b) which show the space-time plot for the variables $x_j^{(1,2)}$ of populations-I and II, respectively.  Figs. \ref{f3}(c,d) show the snapshots of the variables $x_j^{(2)}$ for two different times $t=20$ and $t=35$  (marked by the white solid line in Fig. \ref{f3}(b)) for $\eta=0.15$ and $r=0.1$ after allowing the transients of $5\times 10^5$ iterations.
\begin{figure}[ht!]
\begin{center}
\includegraphics[width=1.1\linewidth]{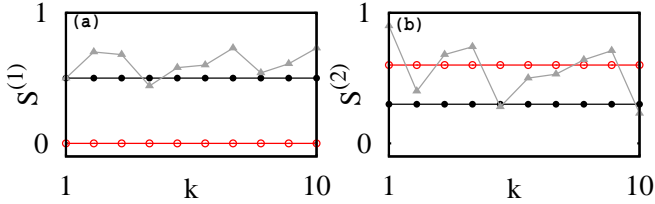}
\end{center}
\vspace{-0.7 cm}
\caption{(Color online) Strength of incoherence plot for three different imperfect synchronized states with 10 bins of time (a) first population  (b) second population where '{\Large $\circ$}' represents imperfect synchronized state, '{\Large $\bullet$}' for mixed imperfect synchronized state, '$\blacktriangle$' for drifted imperfect  synchronized state.  The other parameter values are $c=9$, $r=0.1$, $\sigma=0.1$, $\eta=0.15$, $\omega=1.0$.} 
\label{f3a}
\end{figure} 
\begin{figure}[ht!]
\begin{center}
\includegraphics[width=1.0\linewidth]{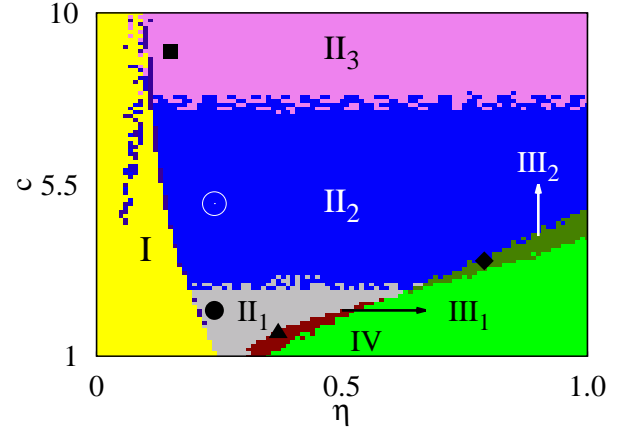}
\end{center}
\vspace{-0.5 cm}
\caption{(Color online) Phase diagram of the system (\ref{g}) in ($\eta,c$) space for $r=0.1$, $\sigma=0.1$, $\omega=1.0$. The regions $I$, $II_1$, $II_2$, $II_3$, $IV$ represent individual synchronization, imperfect synchronized state, mixed imperfect synchronized state, drifted imperfect  synchronized state and globally synchronized state, respectively.  Both the regions $III_1$ (the cluster states are having oscillators within the populations) and $III_2$ (the cluster states are having oscillators from both the populations) represent the cluster synchronized state.  Here `{\large $\bullet$}', `{\Large $\circ$}', `$\blacksquare$', `$\blacktriangle$', '$\blacklozenge$' mark the parameter values corresponding to the Figs. \ref{f1}, \ref{f2}, \ref{f3}, \ref{f5}(a,b) and \ref{f5}(c,d), respectively.}
\label{f4}
\end{figure}

\par By making use of the statistical measure of strength of incoherence introduced by Gopal et al.  \cite{38}, we differentiate the emergence of drifting imperfect synchronized states from other imperfect synchronized states.  For this purpose we divide the total time period $t$ $\in$ (0,T) into $k$ bins ($t_n ,n =1,2, . . . ,k)$ of $t_s$ time units each ($t_s=\frac{T}{k}$).  The strength of incoherence $S$ can be calculated for each time unit and it gives $k$ number of $S$ values.  The strength of incoherence \cite{38} is calculated through the expression
\begin{equation} 
S^{(1,2)}=1-\frac{\sum_{m=1}^{M} s_m^{(1,2)}}{M},s_m^{(1,2)}=\Theta(\delta- \sigma_l(m)^{(1,2)}),
\label{soi}
\end{equation}
where $\delta$ is the threshold value which is small and $\Theta$ is the Heaviside step function. The quantity $\sigma_l(m)^{(1,2)}$, which is the local standard deviation, is calculated from the expression 
\begin{eqnarray} 
\sigma_l(m)^{(1,2)}=\langle(\overline{ \frac{1}{l}\sum_{j=l(m-1)+1}^{ml} \vert w_j^{(1,2)}-\overline{w^{(1,2)}}\vert^2})^{1/2}\rangle_t,\nonumber\\
 m=1,2,...M.
\label{sig}
\end{eqnarray}
for each successive $l$ number of oscillators ($l=N/M$) with $w_j^{(1,2)}=x_j^{(1,2)}-x_{j+1}^{(1,2)}$ and $\overline{w^{(1,2)}}=\frac{1}{N}\sum_{j=1}^Nw_j^{(1,2)}$.  Here $\langle ... \rangle_t$ represents the average over time.  When $\sigma_l(m)^{(1,2)}$ is less than $\delta$, $s_m^{(1,2)}$ takes the value $1$, otherwise it is $0$.  If the imperfect synchronized state is stable $S^{(1,2)}$ yields the same value for all time bins otherwise it varies or differs for different bins, indicating the existence of a drifted imperfect synchronized state.  Figures \ref{f3a}(a,b) are plotted for the strength of incoherence $S^{(1)}$ for population-I and $S^{(2)}$ for population-II, respectively with fixed $\eta=0.15$.  Red line with open circles shows the imperfect synchronized state for $c=2.3$ where $S_1$ takes the value zero (synchronized state in population-I) and $S^{(2)}$ takes the value as constant (solitary state in population-II).  For $c=5$, we can observe both $S^{(1)}$ and $S^{(2)}$ take constant values (black dots) showing the existence of a mixed imperfect synchronized state which is stable with time.  On the other hand for $c=9$ both $S^{(1)}$ and $S^{(2)}$ take different values between zero and one in different bins, and so this figure indicates the unstable nature of the imperfect synchronized state, namely the drifted imperfect synchronized state for the coupling range $r=0.1$.  Here drifting of the solitary oscillators from synchronized group occurs in an irregular manner as a result of the varying value of strength of incoherence as a function of time (or time bins).  Such solitary drift states are closely related to the breathing chimera state as the synchronized group of oscillators exist for certain time period after which it becomes a desynchronized group of oscillators. 
 \par  One can also use the local order parameter \cite{38l}
\begin{equation}
L_j=|\frac{1}{2\delta}\sum_{|j-k|\le \delta} e^{i\theta_k}|,  j=1,2, ...,N
\end{equation}
to measure the degree of (in)coherency which is used to characterize the coherence and incoherence pattern.  Here $\theta_k$ denotes the phase of the $k^{th}$ oscillator.  It is close to unity for the coherent state and decreases in regions of spatial incoherence.  We also note here that characterization of the strength of incoherence is used for systems admitting both phase coherent and nonphase coherent attractors.  Even without introducing the concepts of phase and frequency one can succeed in distinguishing different dynamical states, namely, coherent, incoherent, chimera, multichimera and cluster states in coupled dynamical systems using the concept of strength of incoherence as shown in \cite{38}.

\begin{figure}[ht!]
\begin{center}
\includegraphics[width=1.0\linewidth]{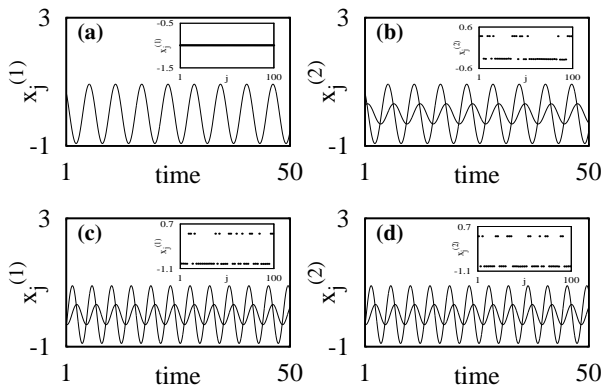}
\end{center}
\vspace{-0.5 cm}
\caption{Time evolution for the variables $x_j^{(1,2)}$ of the cluster states (a) for population-I (b) for population-II with $c=1.5$, $\eta=0.35$ and (c) for population-I (d) population-II with $c=3.3$, $\eta=0.8$.  Other parameter values: $\sigma=0.1$, $\omega=1.0$.  Snapshots for the variables $x_j^{(1,2)}$ of the cluster states are shown in insets of the corresponding figures.}
\label{f5}
\end{figure} 
\begin{figure}[ht!]
\begin{center}
\includegraphics[width=1.0\linewidth]{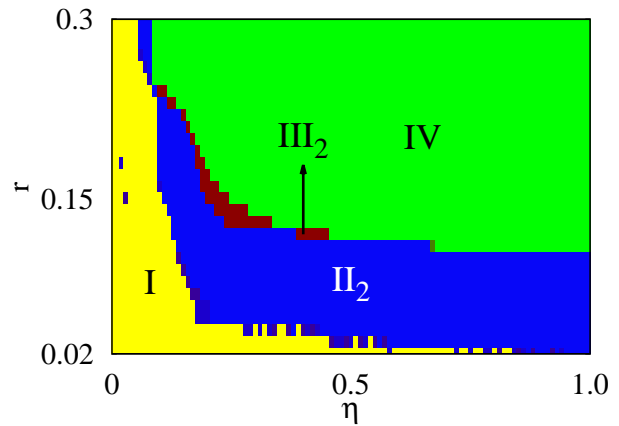}
\end{center}
\vspace{-0.5 cm}
\caption{(Color online) Phase diagram of the system (\ref{g}) in $(\eta,r)$ space for $c=5$, $\sigma=0.1$, $\omega=1.0$. The regions $I$, $II_2$, $III_2$, $IV$ represent individual synchronization, mixed imperfect synchronized state, cluster synchronized state and globally synchronized state, respectively.}
\label{f14}
\end{figure} 
\subsection{Collective dynamics in different parametric spaces}
\par The above studies have been repeated for various values of the coupling strength $\eta$.  To give a global picture of the dynamical states which exist in the two interacting populations of Stuart-Landau oscillators, we have plotted a two phase diagram in the parametric space ($\eta,c$) in Fig. \ref{f4} by fixing $\sigma=0.1$, $\omega=1.0$ for $r=0.1$.  Different dynamical states are identified by making  use of the strength of incoherence ($S$) \cite{38} as described above.  Initially for certain values of $\eta$ the oscillators are individually synchronized with phase difference between the two populations (region-$I$) for all values of the parameter $c$.  On increasing $\eta$ we can observe that the globally synchronized state (region-$IV$) is mediated through the imperfect synchronized states (region-$II_1$) and then through the cluster states (region-$III_1$).  Here a cluster state represents a distinct group of synchronized oscillators having the same amplitude within that group.  Note that for different clusters the amplitudes take different values.  We can observe that in Fig. \ref{f5}(a), all the oscillators in population-I are completely synchronized while the oscillators in population-II split into two groups of synchronized oscillators as shown by the time series plot given in Fig. \ref{f5}(b).  Such synchronized oscillators from both the groups are oscillating periodically and they differ in their amplitudes. Also the members of the oscillators within such a cluster state belong to the same population (as shown in Figs. \ref{f5}(a,b)).  
Then increasing the value of $c$ beyond certain range, the system of oscillators attain complete synchronization via the mixed imperfect synchronized state (region-$II_2$) and then the cluster states (region-$III_2$).  Here the cluster states are having oscillators from both the populations which is different from the above discussed cluster states which is illustrated with the time series plots in Figs. \ref{f5}(c,d) where the snapshots of the variables $x_j^{(1,2)}$ are shown in the insets.  Another interesting phenomenon to be noted here is that the distribution of $x_j^{(1,2)}$ for instantaneous time looks like the oscillation death state (OD) observed in \cite{39s}.  In the case of OD state there is no variation in the distribution of steady states with time while in the case of cluster states one can observe variation in the distribution of state variables because of its oscillating nature with time.
 On further increase in the value of $c$, the mixed imperfect synchronized states become unstable and become drifted imperfect synchronized states (region-$II_3$).  Hence the system of oscillators attains complete synchronization via drifted imperfect synchronized states and then cluster states.  In regions $II_1$, $II_2$ and $II_3$ the synchronized state is unstable, for the reason that small perturbations from synchronized state leads to the onset of imperfect synchronized states, mixed imperfect synchronized states, solitary drift in regions $II_1$, $II_2$ and $III_3$, respectively.  In region-$III_1$ and $III_2$, distribution of initial state near synchronized state leads to complete synchronization and distribution away from the synchronized state leads to the existence of cluster state.
\par To know the robustness of mixed imperfect synchronized states for a wide range of nonlocal coupling, we have plotted the two parameter phase diagram in the ($\eta,r$) parametric space in Fig. \ref{f14}.  In this figure we fix the nonisochronicity parameter as $c=5$ and the coupling strength as $\sigma=0.1$. Initially the oscillators are individually synchronized with phase difference between two populations for small values of coupling strength $\eta$ for all values of the coupling range $(r)$.  For small values of coupling range, one can observe the existence of mixed imperfect synchronized states over a wide range of coupling interaction.  Increase of coupling range to sufficiently larger values of $r$ leads to the suppression of the region corresponding to mixed imperfect synchronized states.  Consequently we cannot observe the presence of mixed imperfect synchronized states when the coupling range approaches the global limit.
\begin{figure}[ht!]
\begin{center}
\includegraphics[width=1.0\linewidth]{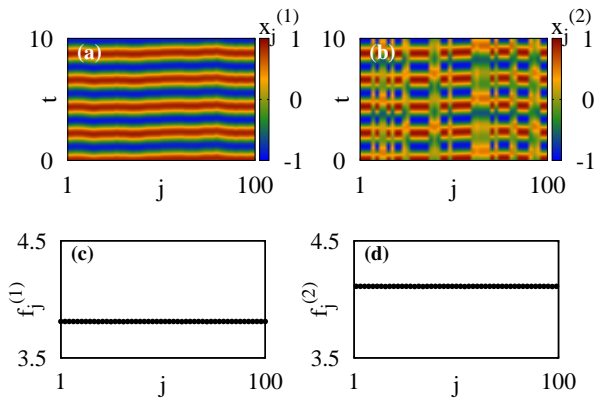}
\end{center}
\vspace{-0.7 cm}
\caption{(Color online) Space-time plots for imperfect synchronized state with frequency mismatch $\varepsilon=0.5$ ($\omega_1=1.0$ and $\omega_2=1.5$) (a) for population-I (b) for population-II with $c=2.3$ and corresponding average frequency of the oscillators in the imperfect synchronized state (c) for population-I and (d) population-II.  Other parameter values: $\sigma=0.1$ and $\eta=0.3.$}
\label{mm1}
\end{figure}
\section{Mixed imperfect synchronized states in the presence of frequency mismatch}
\par  Further we are also interested to investigate the existence of mixed imperfect synchronized states by introducing a frequency mismatch between the two populations such that $\omega_1$ is the frequency of the population-I and $\omega_2=\omega_1+\varepsilon$ is the frequency of the population-II and $\varepsilon$ takes an arbitrary value.  To start with, we first analyze whether the nature of the imperfect synchronized state (which is observed in the absence of frequency mismatch) is robust for an introduction of frequency mismatch between the populations by fixing other parameter values as $r=0.1$, $\sigma=0.1$, $\omega_1=1.0$ and $\omega_2=1.5$.  In the presence of frequency mismatch $\varepsilon=0.5$, for $c=2.3$ we can observe the synchronization in population-I and solitary state in population-II which are illustrated with the space-time plots for the variables $x_j^{(1,2)}$ in Figs. \ref{mm1}(a,b) and the corresponding frequency profiles of the oscillators are shown in Figs. \ref{mm1}(c,d).  Here the oscillators within each population are oscillating with same average frequency. However the synchronized oscillators from both the populations do not share a common average frequency and they differ between the populations as shown in Figs. \ref{mm1}(c,d) (which is distinct from the imperfect synchronized state observed in an identical population where all the oscillators are locked to a common average frequency (Fig. \ref{f1})).  In this case also we choose the synchronized initial state for the oscillators in population-I and uniform initial conditions between $-1$ and $+1$ for the variables $x_j^{(2)}$ and $y_j^{(2)}$ independently. 
\begin{figure}[ht!]
\begin{center}
\includegraphics[width=1.0\linewidth]{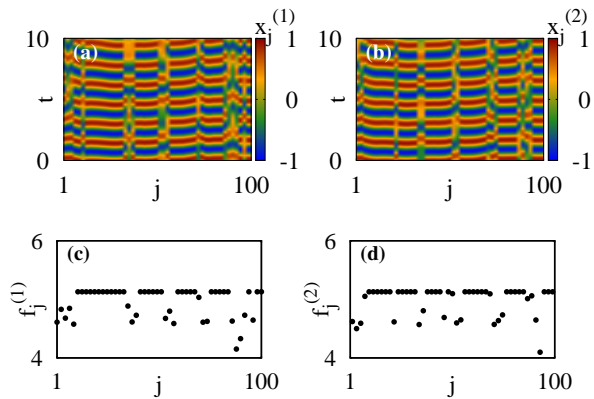}
\end{center}
\vspace{-0.7 cm}
\caption{(Color online) Space-time plots for mixed imperfect synchronized state with frequency mismatch $\varepsilon=0.5$ ($\omega_1=1.0$ and $\omega_2=1.5$) (a) for population-I (b) for population-II with $c=5$ and corresponding average frequency of the oscillators in the mixed imperfect synchronized state (c) for population-I and (d) population-II.  Other parameter values: $\sigma=0.1$ and $\eta=0.3$.}
\label{mm2}
\end{figure}  
\begin{figure}[ht!]
\begin{center}
\includegraphics[width=1.0\linewidth]{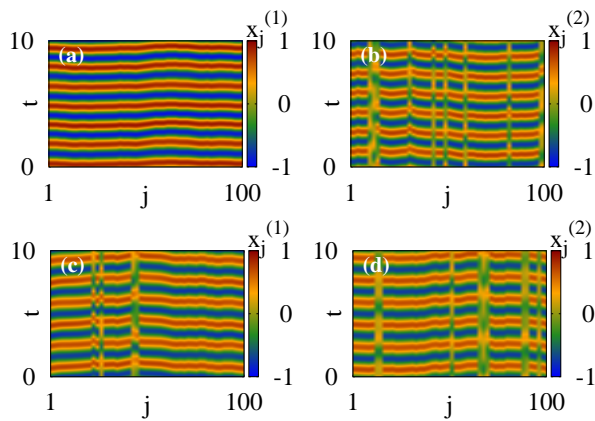}
\end{center}
\vspace{-0.7 cm}
\caption{(Color online) Space time plots for imperfect synchronized state for frequency mismatch $\varepsilon=1.7$ ($\omega_1=1.0$ and $\omega_2=2.7$) (a) for population-I (b) for population-II with $c=5$.  Space time plots for mixed imperfect synchronized state for frequency mismatch $\varepsilon=1.7$ ($\omega_1=1.0$ and $\omega_2=2.7$) (c) for population-I (d) for population-II with $c=6$.}
\label{temc}
\end{figure} 
\par Interestingly, for $c=5$, we can observe the mixed imperfect synchronized states, that is onset of solitary oscillators in both the populations which is illustrated with space time plots of the variables $x_j^{(1,2)}$ in Figs. \ref{mm2} (a, b).  In addition, synchronized oscillators from both the populations do share a common average frequency as shown in Figs. \ref{mm2}(c,d) for $\varepsilon=0.5$ unlike the imperfect synchronized state discussed in Fig. \ref{mm1}.  Further we also analyze the mixed imperfect synchronized states for increasing frequency mismatch between the populations.  For this purpose, we increase the frequency mismatch $\varepsilon$ to $\varepsilon=1.7$ and we investigate the above state with two different values of $c$ for fixed values of $\eta=0.5$, $\omega_1=1.0$ and $r=0.1$.  We can find that for $c=5$ the impact of frequency mismatch leads to the existence of imperfect synchronized state as in Fig. \ref{temc}(a,b).  Interestingly the mixed imperfect synchronized state can be observed for increasing the value of $c$ significantly and Figs. \ref{temc} (c,d) clearly illustrate such state for $c=6.0$.  Thus for significantly small value of nonisochronicity parameter the impact of frequency mismatch between the population dominates the effect of this parameter.  Hence synchronized oscillators in a imperfect synchronized state do not share a common frequency.  If the strength of $c$ is sufficiently large, the effect of nonisochronicity parameter dominates the impact of frequency mismatch.  Thus when the strength of nonisochronicity parameter is sufficiently large it leads to a sharing of common average frequency among the synchronized oscillators in both the populations.
\begin{figure}[ht!]
\begin{center}
\includegraphics[width=1.0\linewidth]{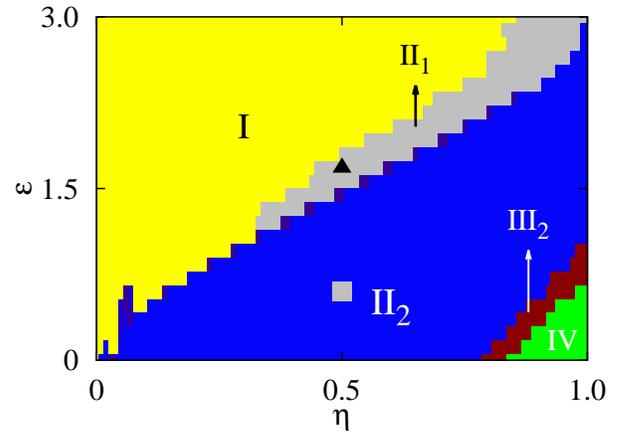}
\end{center}
\vspace{-0.5 cm}
\caption{(Color online) Phase diagram of the system (\ref{g}) in the presence of frequency mismatch $\varepsilon$ ($\omega_1=1.0$ and $\omega_2=\omega_1+\varepsilon$) between the populations for the system parameter values $r=0.1$, $\sigma=0.1$, $c=5$. The regions $I$, $II_1$,$II_2$,$III_2$, $IV$ represent individual synchronization, imperfect synchronized state, mixed imperfect synchronized state, cluster synchronized state and globally synchronized state, respectively. Here `$\blacksquare$' and  `$\blacktriangle$' mark the parameter values corresponding to Figs. (\ref{mm2})(a-d) and (\ref{temc})(a,b), respectively.}
\label{f6}
\end{figure}
\par  We have also plotted the two parameter phase diagram in the parametric space $(\eta, \varepsilon)$ in Fig. \ref{f6} by fixing the frequency of the population-I and varying the frequency of the population-II by $\varepsilon$.  From this, we can find that an increase of frequency mismatch between the populations causes a wide range of individual synchronized region.   This individual synchronization occurs at two different frequencies.    Consequently the regions of mixed imperfect synchronized states start shrinking.  In region- $III_2$, we can observe the cluster states which represent that two different groups having different frequencies as well as two different amplitudes which is different from the cluster state which are observed in the identical case (where two groups having two different amplitudes and frequencies of all the oscillators are the same).

\par   
\section{Conclusion}
\par In summary, we have investigated the existence of different kinds of imperfect synchronized states, including mixed imperfect synchronized states and drifted imperfect synchronized states and chimera states in two interacting populations of nonlocally coupled Stuart-Landau oscillators.  The study shows the existence of synchronized and solitary oscillators where the synchronized group is having the oscillators from both the populations (mixed imperfect synchronized state).  In both imperfect synchronized state and mixed imperfect synchronized state, oscillators from the synchronized group are oscillating periodically whereas the solitary oscillators are quasiperiodic in nature.  We find that when these states are not stable in the sense that, they can drift with time on increasing the value of the nonischronicity parameter.  
\par We have also verified with two parameter phase diagrams that these states can also be observed against the introduction of the frequency mismatch between the two populations.  One can observe the mixed imperfect synchronized state in the presence of mismatch only if the nonisochronicity parameter is sufficiently large to dominate the effect of frequency mismatch between the two populations.  Interestingly, we also find the mixed chimera state for properly chosen initial conditions where we can observe the coexistence of synchronized and desynchronized oscillations namely the mixed chimera state, while synchronized oscillator from both the populations do share a common frequency. 
\section*{Acknowledgements}
 The work of KP forms part of a research project sponsored by NBHM, Government of India.  The work of MS forms part of a research project sponsored by Department of Science and Technology, Government of India.  The work of VKC is supported by the SERB-
DST Fast Track scheme for young scientists under Grant No.YSS/2014/000175.  ML acknowledges the financial support under a NASI Senior Scientist Fellowship program.


\begin{thebibliography}{99}


\bibitem{1}
 A. Pikovsky, M. Rosenblum, and J. Kurths, {\emph Synchronization: A Universal Concept in Nonlinear Science} (Cambridge University Press, Cambridge, 2003).


\bibitem{2}
Y. Kuramoto and D. Battogtokh, Nonlinear Phenom. Complex Syst. {\bf 5}, 380 (2002); D. M. Abrams and S. H. Strogatz, Phys.Rev.Lett. {\bf 93}, 174102 (2004); D. M. Abrams and S. H. Strogatz, Int. J. Bif. Choas {\bf 16(1)}, 21 (2006).


\bibitem{5}
S. I. Shima and Y. Kuramoto, Phys. Rev. E {\bf 69}, 036213 (2004); D. M. Abrams, R. Mirollo, S. H. Strogatz, and D.A.Wiley, Phys. Rev. Lett. {\bf 101}, 084103 (2008).

\bibitem{ma}
 O. E. Omel'chenko, Y. L. Maistrenko, and P. A. Tass Phys. Rev. Lett. {\bf 100}, 044105 (2008); O. E. Omel'chenko, M. Wolfrum, and Y. L. Maistrenko
Phys. Rev. E {\bf 81}, 065201(R) (2010).


\bibitem{7}
G. C. Sethia and A. Sen,  Phys. Rev. Lett. {\bf112}, 144101 (2014);  V. K. Chandrasekar, R. Gopal, A. Venkatesan, and M. Lakshmanan, Phys. Rev. E {\bf 90}, 062913 (2014); K. Premalatha, V. K. Chandrasekar, M. Senthilvelan, and M. Lakshmanan, Phys. Rev. E {\bf 91}, 052915 (2015). 

\bibitem{37a}
C. R. Laing, Phys. Rev. E {\bf 81}, 066221, 2010.
\bibitem{36}
C. R. Laing, Chaos {\bf 19}, 013113, 2009;  C. R. Laing, Physica D, {\bf 238}, 15691588 (2009); C. R. Laing, Chaos {\bf 22}, 043104 (2012).

\bibitem{38b}
E. Montbrio, J. Kurths, and B. Blasius, Phys. Rev. E, {\bf 70}, 056125, (2004); A. Pikovsky and M. Rosenblum, Phys. Rev. Lett. {\bf 101} 264103 (2008); J. H. Sheeba, V. K. Chandrasekar, and M. Lakshmanan, Phys. Rev. E {\bf 79}, 055203(R) (2009); R. Ma, J. Wang, and Z. Liu, Eur. Phys. Lett. {\bf 91}, 40006 (2010); E. A. Martens, M. J. Panaggio, and D. M. Abrams, New J. Phys. {\bf 18}, 022002 (2016).


\bibitem{37t1}
T. Bountis, V. G. Kanas, J. Hizanidis, and A. Bezerianos, Eur. Phys. J. Special topics, {\bf 223}, 721 (2014); M. J. Panaggio, D. M. Abrams, P. Ashwin, and Carlo R. Laing, Phys. Rev. E {\bf 93}, 012218 (2016). 

\bibitem{38m1}
I. Omelchenko, Y. Maistrenko, P. H\"{o}vel, and E. Sch\"{o}ll, Phys. Rev. Lett. {\bf 106}, 234102 (2011).


\bibitem{10}
G. Bordyugov, A. S. Pikovsky, and M. G. Rosenblum, Phys. Rev. E {\bf 82}, 035205 (2010). 

\bibitem{34}
M. R. Tinsley, S. Nkomo, and K. Showalter, Nature Physics {\bf 8}, 662 (2012); S. Nkomo, M. R. Tinsley,  and K. Showalter, Phys. Rev.Lett. {\bf 110}, 244102 (2013).

\bibitem{7a}
A. M. Hagerstrom, T. E. Murphy, R. Roy, P. H\"{o}vel, I. Omel'chenko, E. Sch\"{o}ll, Nature Physics, {\bf 8}, 658 (2012).

\bibitem{34a}
 L. Larger, B. Penkovsky, and Y. Maistrenko, Phys. Rev. Lett. {\bf 111}, 054103 (2013).
\bibitem{33}
E. A. Martens, S. Thutupalli, A. Fourri\`{e}re, and O. Hal-latschek, Proc. Nat. Acad. Sciences {\bf 110}, 10563 (2013).

\bibitem{33a}
C. R. Laing, Phys. Rev. E {\bf 92}, 050904(R) (2015).

\bibitem{33b}
B. K. Bera, D. Ghosh, and M. Lakshmanan, Phys. Rev. E {\bf 93}, 012205 (2016).

\bibitem{27}
N. C. Rattenborg, C. J. Amlaner, S. L. Lima, Neuroscience and Biobehavioral Reviews, {\bf 24}, 817–842 (2000).

\bibitem{27f}
J. M. Davidenko, A. V. Pertsov, R. Salomonsz, W. Baxter, and J. Jalife, Nature {\bf 355}, 349–351 (1992).

\bibitem{28p}
G. Filatrella, A. H. Neilson, and N. F. Pedersen, Eur.Phys. J. B {\bf 61(4)}, 485-491 (2008).

\bibitem{28s}
J. C. Gonzalez, M. G. Cosenza, and M. S. Miguel, Physica A, {\bf 399}, 24, (2014).

\bibitem{28n}
M. Shanahan, Chaos {\bf 20}, 013108, (2010).


\bibitem{Imperfect}
T. Kapitaniak, P. Kuzma, J. Wojewoda, K. Czolczynski, and Y. L. Maistrenko, Scientific Reports {\bf 4}, 6379 (2014). 

\bibitem{39a}
P. Jaros, Y. Maistrenko, and T. Kapitaniak, Phys. Rev. E {\bf 91}, 022907 (2015).

\bibitem{39b}
V. Semenov, A. Zakharova, Y. Maistrenko, and E. Sch\"{o}ll, arXiv1511.03634 (2015).

\bibitem{39c}
A. Zakharova, M. Kapeller, and E. Sch\"{o}ll, Phys. Rev. Lett. {\bf 112}, 154101 (2014). 
 
\bibitem{39s}
I. Schneider, M. Kapeller, S. Loos, A. Zakharova, B. Fiedler, and E. Sch\"{o}ll, Phys. Rev. E {\bf 92}, 052915 (2015).


\bibitem{38}
R. Gopal, V. K. Chandrasekar, A. Venkatesan, and M. Lakshmanan, Phys. Rev. E {\bf 89}, 052914  (2014) 

\bibitem{38l}
I. Omelchenko, Y. Maistrenko, P. Hovel, and E. Sch\"{o}ll, Phys. Rev. L {\bf 106}, 234102 (2011); J. Hizanidis, E. Panagakou, I. Omelchenko, E. Sch\"{o}ll, P. H\"{o}vel, and A. Provata, Phys. Rev. E {\bf 92}, 012915
(2015).
\end{thebibliography}
\end{document}